\documentclass[hidelinks, a4paper,twoside]{article}

\usepackage{comment}
\usepackage{epsfig}
\usepackage{subcaption}
\usepackage{calc}
\usepackage{amssymb}
\usepackage{amstext}
\usepackage{amsmath}
\usepackage{amsthm}
\usepackage{multicol}
\usepackage{pslatex}
\usepackage{apalike}
\usepackage{algorithm2e}
\usepackage[bottom]{footmisc}
\usepackage{xcolor}
\usepackage{svg}
\usepackage{subcaption}
\usepackage{orcidlink}

\usepackage{comment}

\usepackage{SCITEPRESS}     

\begin{document}


\title{Enhancing Holonic Architecture with Natural Language Processing for System of Systems}

\author{\authorname{
Muhammad Ashfaq\sup{1}\,\orcidlink{0000-0003-1870-7680},
Ahmed R. Sadik\sup{2}\,\orcidlink{0000-0001-8291-2211},
Tommi Mikkonen\sup{1}\,\orcidlink{0000-0002-8540-9918},
Muhammad Waseem\sup{1}\,\orcidlink{0000-0001-7488-2577}, and
Niko M\"{a}kitalo\sup{1}\sup{*}\,\orcidlink{0000-0002-7994-3700}}
\affiliation{\sup{1}University of Jyv\"{a}skyl\"{a}, Jyv\"{a}skyl\"{a}, Finland}
\affiliation{\sup{2}Honda Research Institute, Europe}
\email{\sup{*}Corresponding author: niko.k.makitalo@jyu.fi}
}


\abstract{
The complexity and dynamic nature of System of Systems (SoS) necessitate efficient communication mechanisms to ensure interoperability and collaborative functioning among constituent systems, termed holons. 
This paper proposes an innovative approach to enhance holon communication within SoS through the integration of Conversational Generative Intelligence (CGI) techniques. 
Our approach leverages advancements in CGI, specifically Large Language Models (LLMs), to enable holons to understand and act on natural language instructions. This fosters more intuitive human-holon interactions, improving social intelligence and ultimately leading to better coordination among diverse systems.
This position paper outlines a conceptual framework for CGI-enhanced holon interaction, discusses the potential impact on SoS adaptability, usability and efficiency, and sets the stage for future exploration and prototype implementation.
}

\onecolumn \maketitle \normalsize \setcounter{footnote}{0} \vfill

\section{\uppercase{Introduction}}
\label{sec:introduction}


A System of System (SoS) is a collection of systems functioning together to achieve a common goal~\cite{nielsen2015systems}.
These SoSs are comprised of multiple Constituent Systems (CS), each functioning independently with its own management structure. CSs within an SoS can be geographically dispersed, further highlighting the need for robust communication and coordination.
However, when integrated, the overall SoS capabilities are far more than that of the individual CSs forming the SoS. Moreover, a SoS should support evolutionary development allowing CSs to join or leave the SoS at runtime to meet the desired needs. 
Two key SoS instances include sophisticated transport networks and smart energy grids.

Managing the complexity of SoS is a significant challenge. Traditional approaches often struggle due to the inherent autonomy and geographical distribution of Constituent Systems.  Holonic architectures offer a promising solution by decomposing the SoS into smaller, self-governing entities called holons.
This duality facilitates a recursive system architecture, allowing for self-reliance alongside cooperation with other holons forming a holarchy---a hierarchy of holons operating autonomously yet in coordination to achieve common objectives. 
The holonic approach fits best with SoS architectural requirements, motivating researchers to represent the CSs of SoS as \textit{holons} to enable CS discovery and dynamic SoS composition~\cite{elhabbashPrincipled2023}.

While the holonic architecture offers a promising approach for SoS design, significant communication challenges remain.
The heterogeneous CSs, or holons, often adhere to distinct data formats, communication protocols, and interaction patterns. 
This diversity creates interoperability hurdles, hindering seamless information sharing, command interpretation, and task collaboration requiring specialized knowledge for SoS understanding and implementation. 
Additionally, the dynamic nature of SoS, where CSs can join or leave, necessitates adaptive and flexible communication mechanisms to handle evolving SoS compositions. Finally, once operating, SoS might need to interact with humans, necessitating the communication to expand towards a form that is immediately understood by humans.

Equipping holons with Natural Language Processing (NLP) capabilities offers a transformative approach for overcoming  the above communication hurdles. 
Such technologies might enable holons to interpret and respond to natural language instructions, simplifying the holon-to-human interaction and reducing the reliance on specialized knowledge.
Furthermore, this approach facilitates holon-to-holon communication by encoding and decoding machine-executable commands into natural language.  
Consequently, NLP-based communication addresses the underlying heterogeneity of the CSs and their protocols, enhancing overall collaboration and adaptability. 
Recent research explores integrating NLP technologies, especially Large Language Models (LLMs), into robots \cite{koubaa2023rosgpt}, leading to improved human-robot collaboration. 
However, these works lack the multi robot functionality which is necessary for robot-to-robot communication. In addition, the underlying approach is specifically targeted for robots but not for SoS and the Holonic Architecture. 
However, these studies do not incorporate multi-robot functionality, which is essential for robot-to-robot communication. 
Furthermore, their scope is limited to robotics rather than for SoS and Holonic Architecture.

In this paper, we present a conceptual framework that enhances the holonic architecture with NLP capabilities. With advanced NLP technologies, such as LLMs, we enhance holonic architecture for natural language interaction and decision-making in SoS. To the best of our knowledge this is the first work of this nature in the SoS domain.

The rest of this paper is structured as follows. In Section 2, we provide an overview of related work. In Section 3, which forms the core of this paper, we propose a conceptual framework incorporating NLP in holonic architectures. In Section 4, we demonstrate the framework with an Unmanned Vehicle Fleet (UVF) use case scenario. In Section 5, we discuss our findings, and in Section 6, we draw some final conclusions and propose some directions for future work.

\section{\uppercase{Background}}
\label{background}


\subsection{Conversational Generative Intelligence}

Natural Language Processing (NLP) has emerged as a critical field in artificial intelligence, aiming to bridge the gap between human language and computer systems~\cite{khurana2023natural}. NLP encompasses variety of techniques for enabling computers to understand, interpret, and generate human language. 

Within NLP, Large Language Models (LLMs) have revolutionized the ability of machines to process and generate human-like text. 
These LLMs, such as BERT (Bidirectional Encoder Representations from Transformers) and GPT (Generative Pre-trained Transformer), are typically comprised of complex neural networks that are trained on massive datasets of text~\cite{LLMSurvey}. 
These neural networks learn complex language features, such as patterns, structures, context, and semantics, allowing them to provide advanced NLP capabilities such as  text classification, sentiment analysis, translation, and question-answering. 

A specific application of NLP and LLM technologies lie in Conversational Generative Artificial Intelligence (CGI).
By leveraging conversational and generative intelligence technologies, CGI aims to create systems that can engage in natural and meaningful conversations with humans \cite{dwivedi2023so}.
CGI has applications in robots, virtual assistants, and other interactive systems designed to simulate natural conversations.

\subsection{Holonic Architecture for System of Systems}

Holons are autonomous and integrally connected entities that exhibit independent functionalities while contributing to a larger system~\cite{koestler1968ghost}.
Thanks to their dual nature, holons are excellent for modeling the heterogeneous CSs to properly reflecting their independent functions and contributions to the overall SoS \cite{blair2015holons}. 

This approach has been further extended by \textit{holonic architecture} that supports  dynamic SoS composition~\cite{nundloll2020ontological}, CS discovery, and ad-hoc scalability~\cite{sadik2023self}.
\cite{elhabbashPrincipled2023} implemented the holonic architecture using an ontology-based approach, enabling CSs to reason and understand each other, facilitating dynamic SoS composition. 
However, this work assumes that the ontological descriptions of the holons are manually provided by vendors or systems engineers.
Addressing this limitation, \cite{zhang2023nlp} proposed an NLP-based approach that automatically extracts ontological description of IoT devices by scraping web data.
While this approach offers automation, the architecture currently lacks capabilities for holon-to-environment communication~\cite{halba2021framework}, human-to-holon interaction, and communication with unknown holons.

\section{\uppercase{Conceptual Framework}}

\begin{figure*}[!ht]
  \centering
  \includegraphics[width=\textwidth]{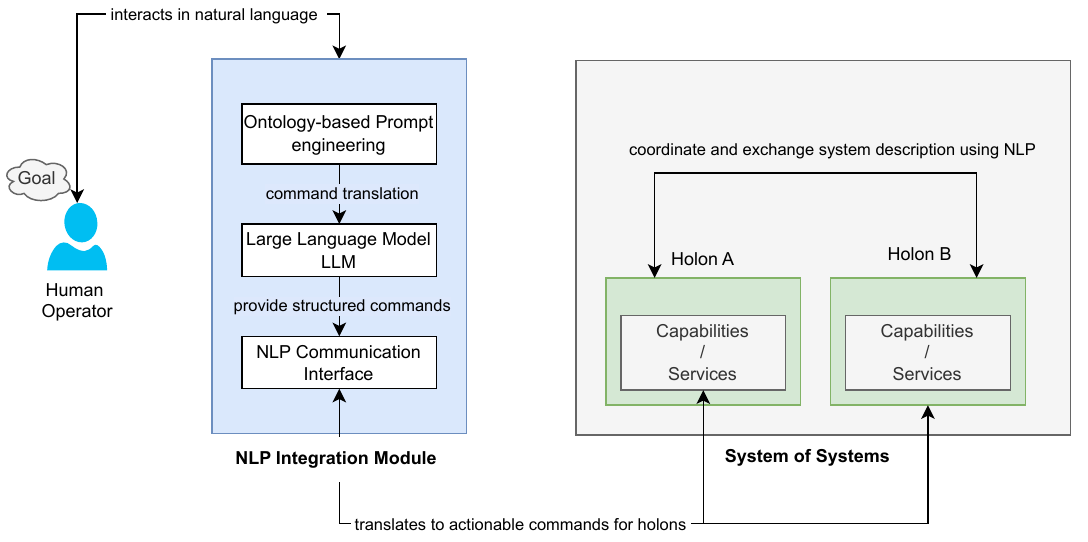}
  \caption{The conceptual framework showing NLP integration with the Holonic Architecture}
  \label{fig:concept}
 \end{figure*}

Figure~\ref{fig:concept} outlines the NLP-enhanced holon communication model, depicting its components and their interactions.
Among these components is the human operator who provides natural language instructions to the holons. For instance, the operator can broadcast a high-level goal they envision. This goal usually requires the collaboration of multiple holons, exceeding the capabilities of a single entity. Consequently, the human operator must identify a group of holons capable of collaboratively achieving this goal. 
As depicted, each holon possesses specific capabilities or services, which are the resources integrated within them.
 These capabilities can be offered by the holons to contribute to achieving the operator's overarching goal.

At the heart of the framework lies the \textit{NLP module}, serving as the bridge between the human operator and the holons.  This module comprises three components: \textit{ontology-based prompt engineering}, a \textit{LLM}, and a \textit{NLP communication}. The \textit{ontology-based prompt engineering component} refines the human operator's input with relevant context. This input can be provided in various formats, including text, speech, or visuals.  One approach of incorporating context involves incorporating domain-specific ontology information into the prompt, as demonstrated in \cite{koubaa2023rosgpt}. This step is crucial to prevent the LLM from generating overly generic outputs that lack domain applicability.
The \textit{LLM component} leverages NLP capabilities for understanding and generating natural language interactions. 
It interprets the input, processes it, and converts it into a syntax comprehensible to the holons. The resulting structured command is then passed to the \textit{NLP communication component}. This interface translates the processed instructions into actionable commands that can activate or deactivate the holons' capabilities or services as needed. Notably, NLP module can also facilitate coordination among holons, enabling them to exchange descriptions of their encapsulated systems.

For example, suppose the human operator aims to achieve a specific goal. Traditionally, achieving such a goal would require the operator to be familiar with the internal workings of each CS. The operator would then need to orchestrate these CSs together to design an SoS that accomplishes the goal. 
In our proposed framework, the operator communicates their goals to the NLP module. This module then liaises with the available holons regarding the goal's requirements.  
The holons iteratively engage with the module for further clarifications, if needed. Following this negotiation process, certain holons commit to providing specific services, while others may be deemed irrelevant to the goal.
The relevant holons then form a \textit{holon composition}, collaborating with each other to fulfill the given mission.

\section{\uppercase{Use Case Scenario}}
\label{sec:case_study}

\begin{figure*}[ht!]
    \centering
    \begin{subfigure}[t]{0.5\textwidth}
        \centering
        \includegraphics{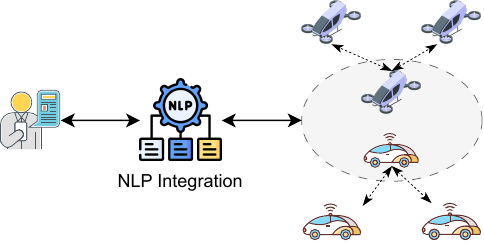}
        \caption{Human-UV Interaction and Planning}
    \end{subfigure}%
    ~ 
    \begin{subfigure}[t]{0.5\textwidth}
        \centering
        \includegraphics{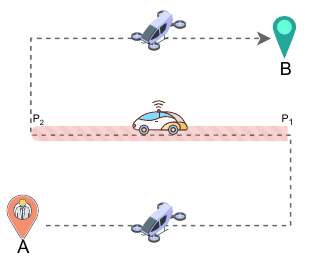}
        \caption{Resulting UVF of two UAVs and one UGV (P\textsubscript{1}---P\textsubscript{2} is no-fly zone)}
    \end{subfigure}
      \vspace{10pt} 
    \caption{Caption place holder}
    \label{fig:case_study}
\end{figure*}

\subsection{Overview}

This case study explores the integration of the proposed framework into the smart city transportation using Unmanned Vehicle Fleets (UVF)---a SoS that demonstrates complex interaction dynamics and scalability challenges. 

The UVF, characterized by its ad-hoc scalability and dynamic environment, faces challenges such as changing missions, increasing range and capacity requirements, UV failures, and battery constraints~\cite{sadik2023self}. 
Within this SoS, the UVF is composed of numerous unmanned entities, including Unmanned Ground Vehicles (UGVs) and Unmanned Aerial Vehicles (UAVs), each functioning autonomously as a holon and concurrently in concert with the entire fleet~\cite{tchappi2020critical}, exemplifying the quintessential principles of holonic systems in managing complex SoS environments.

\subsection{Sample Scenario}

Consider an urban mobility scenario where a resident needs to move from Position A to Position B in a smart city. This scenario involves quickly and efficiently navigating the complex cityscape with some no-fly zones, and traffic and road conditions. This includes following challenges:

\begin{itemize}
    \item \textit{Complex Urban Environment}: Navigating through a densely populated urban area with varying altitudes and no-fly zones for UAVs.
    \item \textit{Dynamic Routing}: Adapting to real-time traffic and environmental conditions to ensure the fastest and safest route.
    \item \textit{Vehicle Coordination}: Seamlessly transitioning between UAVs and UGVs while maintaining a consistent and comfortable experience.
    \item \textit{Communication}: Ensuring clear and efficient communication between the user, UVs, and the control center to manage expectations and adapt to any changes in the mission.
\end{itemize}

\subsection{NLP-enhanced UVF Communication Framework}

The user opts to use the city's UVF service, which leverages a mix of UAVs and UGVs to transport him to his destination. 
Figure \ref{fig:case_study} shows the integration of NLP capabilities into the UVF communication framework.
This integration enables a twofold enhancement: it streamlines human-to-fleet interactions and automates intra-fleet communications.
The process involves following steps:

As shown in Figure \ref{fig:case_study}a, the user communicates his destination through a simple natural language interface to the urban mobility service. The service uses NLP to interpret the user's request and initiates the mission planning process. To optimize communication efficiency, one UV from both UGV and UAV swarm is chosen as a representative from the swarm. This representative UV interacts with the user and the urban mobility service.  Meanwhile, the other UVs exchange information and synchronize their actions with the representative UV. In this way, all UVs negotiate roles, paths, and timing based on their capabilities, current status, and environmental factors. 

After the negotiation process the user and UVs learn that the path from Position A to Position B includes no flying zone (represented by P\textsubscript{1}---P\textsubscript{2} in Figure \ref{fig:case_study}b). 
Considering this, a UVF is formed that comprises of two UAVs for aerial segments and one UGV for ground transportation.
Upon reaching the predetermined landing zone close to user's location (Position A), the first UAV covers the  communicates with the UGV to prepare for a smooth transition for no-fly zone. The user is then transported by the UGV to a launch zone closer to his destination, where the second UAV takes over to complete the final leg of the journey.

\section{\uppercase{Conclusion and Future Work}}
This paper introduces a novel approach to enhance holon communication within SoS through Natural Language Processing, aiming to bridge the communication gap between human operators and holons, and among holons themselves. The proposed framework showcase the potential for NLP to significantly improve the efficiency, adaptability, and usability of SoS, paving the way for more intuitive and effective system-level collaboration.

In general, the field represents a promising area of ongoing research, with future developments anticipated to further refine and validate the proposed model in practical SoS applications. In short term, building on experiences with the ROS2 platform (anonymous, xxxx)
the implementation of the proposed framework is currently underway, using 
[omitted for blind review] (anonymous, xxx)
as the underlying technology. 
 



\bibliographystyle{apalike}
{\small
\bibliography{main}}

\end{document}